# Relative Spectral Response Function Retrieval of Hyperspectral Instruments in Atmospheric Spectrometry

P. Dussarrat, G. Deschamps
EUMETSAT, Eumetsat allee 1, 64285 Darmstadt, Germany
pierre.dussarrat@eumetsat.int

*Abstract*— In the last 20 years, remote sensing from space has become essential to monitor the Earth's surface and atmosphere. Nowadays, tens of satellites carry hyperspectral spectrometers operating from the infrared to the ultraviolet. Such instruments allow decomposing the light that exits the atmosphere from its top into hundreds to thousands of contiguous spectral channels. By analysis of the light spectral distribution, and in particular the depths of selected absorption lines, researchers and meteorological agencies can retrieve the atmosphere composition and thermodynamic state. To get a global view of the Earth, several instruments are generally operated synergistically, therefore, a harmonized calibration must be achieved between them.

To cross-calibrate two spectrometers, a common practice is to analyze an ensemble of collocated measurements [1], meaning acquisitions performed at the same time and under the same geometry. Nonetheless, such analysis always faces the issue of setting appropriate temporal and geometric thresholds in defining the collocations, trading off between statistics and quality. Consequently, some collocation mismatches may have a substantial impact on the cross-calibration results. Thus, the following manuscript describes in detail the inclusion of collocation errors into the mathematical description and presents an application which is designed on purpose to be robust to such errors.

Then, the knowledge of the spectral sensitivities of each channel to the incoming light, called the spectral response functions (SRF), are key to the exploitation of the acquisitions. In particular, the exact shapes of the recorded absorption lines are directly related to the SRF. In that context, the authors have studied and designed a novel methodology to retrieve relative SRF between two or more spectrometers, within a single instrument or between instruments embarked on different platforms. The objective of the methodology is to characterize discrepancies of responses between flying spectrometers, track long-term evolutions and harmonize their responses with post-processing when necessary.

## I. Introduction

The following manuscript introduces a novel methodology to compute a relative SRF between two hyperspectral spectrometers by analysis of a large dataset of collocated acquisitions taken from the orbit. The methodology is aimed at any type of spectrometer as we only assume that the two instruments acquire spectra composed of contiguous channels spanning the same spectral band.

The first objective is to design a mathematical solution allowing the retrieval of a relative SRF between the two spectrometers accurately enough so that it can be interpreted in terms of instrumental or processing limitations and defects. As an example, that activity may be adapted into the monitoring routines aimed at tracking launch-induced or in-flight long-term degradations of the instruments. The second objective is to use the derived relative SRF to perform a harmonization of the acquisitions, as harmonized data is essential to provide a coherent view of the Earth's system.

As discussed thoroughly in the following, cross-calibrations may be significantly impacted by the limited quality of the collocations, therefore the solution presented in this manuscript is designed on purpose to be robust to some extent to collocation mismatches. By doing so, we will see that it enables the comparison of spectrometers whose footprints are not even spatially overlapping, which opens largely the scope of applications.

This work is in the continuity of a previous study realized in the context of the relative SRF retrieval of the Metop-IASI instruments [2]. In the following, we present an improved mathematical description and a more general and efficient solution. To the authors' knowledge, it is the first time that the impact of collocation quality is examined mathematically in such a way in the field of remote sensing.

Section II presents the hypothesis and a first look upon the problem of retrieving relative SRF. Section III presents the theoretical background leading to the design of a mathematical solution retrieving accurately the relative SRF, with the special property of being robust to some extent to collocation errors and radiometric noise. Section IV discusses how to extend the methodology to more than two spectrometers and apply it to imagers.

## II. First Look

The following section provides a first look into the problem of retrieving a relative SRF between two hyperspectral spectrometers and introduces a mathematical description of the collocation quality.

### A. Hypothesis

Let's consider a dataset composed by $N_a$ collocated acquisitions of two hyperspectral spectrometers acquiring continuous spectra over the same spectral band. The two sets of spectra are noted $(X,Y)$, their sizes are respectively $(N_x,N_a)$ and $(N_y,N_a)$, with $N_x$ and $N_y$ being their respective number of channels. In the following, we will assume that $N_x \geq N_y$, if it is not the case one can always swap both sets.

The basic idea of this work is to retrieve a so-called relative SRF matrix $g$ of size $(N_y, N_x)$, that best relates the sets $X$ and $Y$ as follows:

$$Y = g X \qquad (II.A1)$$

It is important to note that, in the following we always assume implicitly that the differences between both sets are uniquely caused by a difference of SRF (exclusion of non-linearity, straylight, shapes of footprints differences…), and that the offsets of the two instruments are well calibrated. Moreover, we stress that $g$ is generally not unique (existence of a null space), and the following derivation only provides one solution.

We want to write the solution in terms of variance and covariance matrices, $cov[A,B] = \langle AB^T \rangle$ and $var[A] = cov[A,A] = \langle AA^T \rangle$, evaluated over the whole dataset, where "T" stands for the transpose operator. Variance and covariance matrices have the advantage of being generally much smaller than the acquisition sets and they can be computed iteratively (by splitting the dataset).

This is solved by basic linear regression and the solution, called in the following first guess and noted $g_1$, writes:

$$g_1 = cov[Y,X] \, var[X]^{-1} \qquad (II.A2)$$

Where we introduce the inverse of the covariance matrix of X, which generally always exists (in practice because of the presence of radiometric noise).

Nonetheless, we will prove in the following that this solution is sensitive to the collocation quality of the acquisitions. Therefore, the relative SRF is generally biased, and the solution must be improved.

### B. Collocation errors

In this section, we introduce collocation errors into the mathematical model. We assume that set $X$, is impacted by collocation shifts, spatial and/or temporal, whose radiometric effects on the spectra are noted $\delta$, of size $(N_x, N_a)$, which are mostly unknown. We assume nonetheless that the collocation errors are small compared to the intrinsic variability of $X$, that is to say: $var[\delta] \, var[X]^{-1} \ll I_x$, with $I_x$, the identity matrix of size $(N_x, N_x)$. It also implies that the collocation errors transposed into the $Y$ dataset, noted $\Delta = g \, \delta$, are small compared to the variability of $Y$: $var[\Delta] \, var[Y]^{-1} \ll I_y$, with $I_y$, the identity matrix of size $(N_y, N_y)$.

We note the set free of collocation error $x$ and therefore the acquisitions $X$ are composed of the sum of $x$ and $\delta$:

$$X = x + \delta \qquad (II.B1)$$

In the following, we will look for a new solution for the relative SRF that aims at waiving the impact of the collocation errors on the retrieval. Therefore, $g$ must realize:

$$Y = g x \qquad (II.B2)$$

The objectives of the following sections are first to prove that $g$ can be approached very accurately even in the presence of collocation errors and radiometric noise by $g'$ that writes:

$$g' = (cov[Y,X] \, var[X]^{-1} \, cov[X,Y] \, var[Y]^{-1})^{-1/2} \qquad (II.B3)$$
$$cov[Y,X] \, var[X]^{-1}$$

The second objective is to show that the set $X$ can be post-processed into "Y-like" data by performing $Y' = g' \, X$ that compensates effectively the difference of SRF of $X$ with respect to $Y$.

### III. THEORETICAL BACKGROUND

The following sections present step by step the assumptions and calculations leading to the proof that the solution $g'$ (II.B3) is a very accurate evaluation of $g$ even in presence of collocation errors and radiometric noise, and can be used to post-process dataset $X$.

### A. Collocation error properties

To be able to solve this problem we must first set constraints on the collocation errors statistical properties. By assuming that the dataset spans a large diversity of scenes, we can postulate that the variances of the sets with and without collocation errors are the same: $var[x] = var[X]$ and that there is a symmetry of their covariance: $cov[X,x] = cov[x,X]$. Therefore, by simple manipulation of these two equations we find a relation for the covariance of $x$ and $\delta$ as function of the variance of $\delta$:

$$cov[\delta,x] = cov[x,\delta] = - var[\delta]/2 \qquad (III.A1)$$

In the following, we take advantage of such relation to evaluate several solutions for the relative SRF and their sensitivity to the collocation errors.

### B. Evaluation of var[Y]

We start by evaluating an intermediate result about the covariance of $Y$ as function of the covariance of $X$. By making use of the equations (II.B1-2), we find directly that:

$$var[Y] = g \, var[X] \, g^T \qquad (III.B1)$$

Such results show that if we can evaluate $g$, then, the post-processed dataset $Y' = g \, X$ would naturally exhibit the same variance as $Y$.

### C. Evaluation of the first guess $g_1$

By using the newly introduced formalism accounting for collocation errors, we can evaluate the accuracy of the first guess solution $g_1$, as presented in section II.A, now in the presence of collocation errors. We find that:

$$cov[Y,X] \, var[X]^{-1} = g - g/2 \, var[\delta] \, var[X]^{-1} \qquad (III.C1)$$

Consequently, $g_1 = cov[Y,X]\ var[X]^{-1}$, is a rather good approximation of $g$, but it is biased at first order in collocation errors ($var[\delta]\ var[X]^{-1}$). Therefore, we look for a more accurate solution.

### D. Evaluation of the right-inverse first guess $g_2^{-1}$

By re-working the previous equations, one can find the following relation:

$$g\ cov[X,Y]\ var[Y]^{-1} = I_y - \tfrac{1}{2}\ var[\Delta]\ var[Y]^{-1} \quad \text{(III.D1)}$$

It reveals a rather good evaluation of the right-inverse of $g$, that we note $g_2^{-1} = cov[X,Y]\ var[Y]^{-1}$. But, once again it is biased at first order in the collocation errors ($var[\Delta]\ var[Y]^{-1}$).

Note that, in the general case, $g$ is not invertible. A particular and useful case, in which the inverse exists is discussed in section III.F.

### E. Evaluation of $g_1 g_2^{-1}$

To compute $g'$, we need first to evaluate the intermediate result $g_1 g_2^{-1}$. By mixing equations III.C1 and III.D1, we find by retaining only the terms at first order in the collocation errors:

$$g_1 g_2^{-1} = cov[Y,X]\ var[X]^{-1}\ cov[X,Y]\ var[Y]^{-1} \quad \text{(III.E1)}$$
$$= I_y - var[\Delta]\ var[Y]^{-1} + o\{var[\Delta]\ var[Y]^{-1}\}$$

Where we introduce the notation, $o\{var[\Delta]\ var[Y]^{-1}\}$, using the "little-o", meaning that the residual is negligible compared to $var[\Delta]\ var[Y]^{-1}$ which is already assumed small, thus we say that the residual is at second order in collocation errors. Then, we can also evaluate its inverse square root at first order in the collocation errors:

$$(g_1 g_2^{-1})^{-1/2} \quad \text{(III.E2)}$$
$$= I_y + \tfrac{1}{2}\ var[\Delta]\ var[Y]^{-1} + o\{var[\Delta]\ var[Y]^{-1}\}$$

We witness that $g_1$ and $g_2^{-1}$ do not cancel each other, with a residual at the level of $var[\Delta]\ var[Y]^{-1}$.

### F. Evaluation of $g'$

To compute $g'$, we mix equations (III.E2) and (III.C1), and we keep the terms at first order in the collocation errors:

$$g' = (g_1 g_2^{-1})^{-1/2}\ g_1 \quad \text{(III.F1)}$$
$$= g + \tfrac{1}{2}\ (var[\Delta]\ var[Y]^{-1} g - g\ var[\delta]\ var[X]^{-1})$$
$$+ o\{g\ var[\delta]\ var[X]^{-1}\}$$

Therefore, we show that $g'$ is a very good approximation of $g$, even if the residual remains at first order in $var[\delta]\ var[X]^{-1}$ and $var[\Delta]\ var[Y]^{-1}$. But as the terms within the parenthesis tend to cancel each other, $g'$ is generally more accurate than the first guess $g_1$.

The evaluation becomes very accurate, at second order in the collocation errors, when the inverse of $g$ can be defined. To do so, it is necessary but not sufficient that both spectrometers possess the same number of channels, $N_x = N_y$. Thus, equation (III.B1) can be re-written as $var[X]^{-1} = g^T\ var[Y]^{-1} g$ and equation (III.F1) becomes:

$$g' = g \times (I + o\{var[\delta]\ var[X]^{-1}\}) \quad \text{(III.F2)}$$

This case is the most useful one as the retrieval becomes accurate at second order in the collocation errors. Therefore, it allows interpreting $g'$ in terms of instrumental or processing defects waiving the potential impact of collocation mismatches. It is realized, for example, when comparing two similar spectrometers that have the same exact spectral channels and for which the relative SRF is close to the identity matrix, as discussed in more detail in section (IV.A).

### G. Post-processing

Finally, we verify that $g'$ can be used to post-process the set $X$ to transform it into a "$Y$-like" set by compensating the difference in relative SRF. Note that it is done without any assumptions about the inversibility of $g$, as made at the end of the previous section.

Indeed, by computing the variance of the post-processed set $Y' = g'\ X$ and keeping the terms at first order in collocation error, we find that its variance is very close to the variance of $Y$:

$$var[Y']\ var[Y]^{-1} = I_y + o\{var[\Delta]\ var[Y]^{-1}\} \quad \text{(III.G1)}$$

Moreover, using equation (III.B1), we find that applying $g'$ is extremely close to applying $g$ in the post-processing:

$$var[(g - g')\ X]\ var[Y]^{-1} = o\{var[\Delta]\ var[Y]^{-1}\}) \quad \text{(III.G2)}$$

Therefore, post-processing is always expected to be very effective as the residuals are at second order in the collocation errors ($var[\Delta]\ var[Y]^{-1}$), even if the inverse of $g$ does not exist.

### H. Radiometric noise

Finally, the previous demonstration can be adapted to account for the presence of radiometric noise for both spectrometers. We remove the collocation errors and introduce random noises to both sets, noted ($n_x$, $n_y$), uncorrelated to all other components and whose variances are assumed to be unknown. We assume nonetheless that the noises variances are small compared to the intrinsic variabilities of the sets, $var[n_x]\ var[X]^{-1} \ll I_x$ and $var[n_y]\ var[Y]^{-1} \ll I_y$. The sets write:

$$X = x + n_x\ ;\ Y = y + n_y \quad \text{(III.H1)}$$

Therefore, we look for a solution $g$ that waives the effect of noise on the dataset:

$$y = g\ x \quad \text{(III.H2)}$$

Then, equations (III.C1 and D1) become:

$$g_1 = g - g\, var[n_x]\, var[X]^{-1} \quad (III.H3)$$

$$g\, g_2^{-1} = I_y - var[n_y]\, var[Y]^{-1} \quad (III.H4)$$

We witness that, as for the collocation errors, the radiometric noises produce biases to the first guess $g_1$ and right-inverse first guess $g_2^{-1}$ at first order in radiometric noise ($var[n_x]\, var[X]^{-1}$ and $var[n_y]\, var[Y]^{-1}$). Then, we can evaluate $g_1 g_2^{-1}$:

$$g_1 g_2^{-1} = \quad (III.H5)$$
$$= I_y - (var[n_y]\, var[Y]^{-1} + g\, var[n_x]\, var[X]^{-1}\, g_2^{-1})$$

And finally, $g'$, keeping only the terms at first order in the radiometric noises:

$$g' = g \quad (III.H6)$$
$$+ \tfrac{1}{2}\,(var[n_y]\, var[Y]^{-1}\, g + g\, var[n_x]\, var[X]^{-1}(g_2^{-1}g - 2\, I_x))$$
$$+ o\{g\, var[n_x]\, var[X]^{-1}\}$$

Once again, the result is biased at first order in the radiometric noise, but the terms in the parenthesis tend to cancel each other. And if we assume that the inverse of $g$ exists, then $g_2^{-1} = g^{-1}$, and:

$$g' = g \quad (III.H7)$$
$$+ \tfrac{1}{2}\,(var[n_y]\, var[Y]^{-1}\, g - g\, var[n_x]\, var[X]^{-1})$$
$$+ o\{g\, var[n_x]\, var[X]^{-1}\}$$

Moreover, if the noises of the two spectrometers are equivalent $var[n_y] = g\, var[n_x]\, g^T$, then using $var[X]^{-1} = g^T\, var[Y]^{-1}\, g$, the evaluation becomes accurate at second order in radiometric noise:

$$g' = g\,(I + o\{var[n_x]\, var[X]^{-1}\}) \quad (III.H8)$$

The post-correction can also be shown to be accurate at second order in the radiometric noise. Finally, the result may be extended to a case in which both small radiometric noise and collocation errors are present at the same time.

To conclude section III, we have demonstrated that the solution $g'$ (II.B3) is robust to the presence of collocation errors and radiometric noise, and can be used to post-process the set $X$. We underline that $g'$ (II.B3) is always better suited than the basic first guess $g_1$ (II.A2).

## IV. APPLICATIONS

The following section presents several extensions of the methodology, discusses limitations and presents possible improvements.

### A. Several spectrometers within an instrument

We discuss in the following how to deal with an instrument concatenating many hyperspectral spectrometers.

For example, an instrument based on a grating generally exploits one dimension of the focal plane for the dispersion of the light and the other for spatial coverage. Therefore, the instrument is composed of several spatially concatenated spectrometers, with each spectrometer acquiring spectra over different spatial aera along a line of about 10 to 100 km with small to no footprint overlapping between them. We can still consider that the different spectrometers perform simultaneous acquisitions that are, to some extent, always spatially collocated as they span a limited area of the Earth's disk. Indeed, the atmosphere and surface generally exhibit large spatial correlation at this scale. Thus, we can build easily large datasets of rather well-collocated acquisitions between the different spectrometers.

Consequently, we can apply the previous results by retrieving a relative SRF between all pairs of spectrometers. This is made possible by the fact that the solution was made robust on purpose to the presence of collocation errors.

This situation is the most favourable as all spectrometers have generally the same spectral channels, a similar noise and the relative SRF is expected close to the identity matrix. Therefore, we are in conditions in which the inverse of $g$ exists and therefore the relative SRF retrieval is expected to be accurate at second order in the collocation errors and radiometric noise (see discussion in section III.F-H).

By noting the sets associated with the $N$ spectrometers $X_{i=1..N}$, the relative SRF between spectrometers $i$ and $j$, noted $g'_{i,j}$, writes:

$$g'_{j,i} = (cov[X_j,X_i]\, var[X_i]^{-1}\, cov[X_i,X_j]\, var[X_j]^{-1})^{-1/2} \quad (IV.A1)$$
$$cov[X_j,X_i]\, var[X_i]^{-1}$$

Then the different relative SRF can be assembled into one single relative SRF per spectrometer with respect to all the others, noted $g'_{\#i}$, by average of (IV.A1):

$$g'_{\#i} = 1/(N-1)\, \sum_{j \neq i} g'_{j,i} \quad (IV.A2)$$

Note that it is important to perform the retrieval of relative SRF per pair of spectrometers first and then merge them. Averaging the acquisitions of all spectrometers and then perform a retrieval for each of them with respect to the average would yield wrong results as the method requires comparing acquisitions with similar footprints.

Such result can finally be used to harmonize the acquisitions by computing: $X_i' = g'_{\#i}\, X_i$. It is expected to remove effectively potential discrepancies of responses between spectrometers.

Similarly, the methodology can be applied to assess relative SRF between similar spectrometers embarked on different platforms (see the analysis of two distinct IASI instruments in reference [2]).

## B. Mono-channel spectrometers: "Imagers"

The equations can be simplified in the case of a "imager" instrument acquiring distinct spectral channels. In that case, each channel can be analyzed separately, the variance and covariance become scalars, as well as the relative SRF, which transforms into a relative gain between sets $X$ and $Y$:

$$g' = (var[X] / var[Y])^{1/2} \quad \text{(IV.B1)}$$

The solution becomes extremely simple, it is a basic renormalization of standard deviation of the two sets. But the relative gain assessment retains the same properties as the general result, it is robust to the presence of collocation errors and radiometric noise.

Following section (IV.A), the method can be extended to equalize the gain of $N$ detectors within an imager instrument, by computing a single corrective gain per detector with respect to all the others:

$$g'_{\#i} = 1/(N-1) \sum_{j \neq i} (var[X_j] / var[X_i])^{1/2} \quad \text{(IV.B2)}$$

As an example, such methodology was used successfully to remove stripes and calibration artefacts of the infrared channels of the MTG-I1 FCI instrument, by using standard daily acquisitions and characterizing each detector with respect to all the others (as reported at the GSICS meeting 2025 [3]).

## C. Limited information

The methodology can face critical stability issues for the evaluation of $g$ (extreme sensitivity to the inputs) when dealing with datasets with limited diversity of scenes. That is caused by the fact that there are not enough independent pieces of information to define a relative SRF.

When comparing similar spectrometers ($N_x = N_y$), if we also assume that the relative SRF is mostly achromatic (rather constant diagonals) and local (negligible term far from the central diagonal), we can adjust the calculations to increase artificially the diversity of acquisitions which stabilizes the retrieval. To do so, we spectrally slice all acquisitions into many smaller versions of sizes $N_{x,sub}$. For each acquired spectra, it exists ($N_x - N_{x,sub} + 1$) of such sliced spectra. Then, the problem becomes retrieving a smaller version of the relative SRF, of size ($N_{x,sub}, N_{x,sub}$), relating now all sliced spectra. Finally, it is a matter of trading-off between the extension of the relative SRF matrix to retrieve and the stability of the retrieval by adjusting $N_{x,sub}$.

In practice, such operation can be realized equivalently by extracting and averaging all square matrices of size ($N_{x,sub}$, $N_{x,sub}$) along the main diagonal of the different variance and covariance matrices present in equation (II.B3). Then, the retrieval is performed with the same equation but applied to the smaller versions of the variance and covariance matrices (see example in section 2.6 of reference [2], in which the method is named "subsetting"). Finally, the full SRF matrix is evaluated approximately by duplication of the central row of the retrieved smaller SRF matrix, stacked diagonally to form the full SRF.

## D. Uncertainty evaluation

To evaluate the uncertainty of the relative SRF retrieval, several complementary evaluations may be performed:

Residual: As discussed throughout the article, for the methodology to remain accurate, the noise and collocation errors should be small. Therefore, one can first evaluate $g'$ and then verify a posteriori the following relation: $var[Y - g'X] \, var[Y]^{-1} \ll I_y$ (or its sliced version, as introduced in IV.C). In general, one should verify that the difference is smaller than 10%.

Statistical dispersion: The retrieval can be sensitive to the limited statistics of the collocation dataset. To evaluate this sensitivity, one can perform a bootstrapping of the dataset [4]. We compute several times the relative SRF after performing random picks with replacement inside the initial dataset of collocations. Finally, we evaluate the statistical dispersion of the results by computation of standard deviations.

Consistency check: Finally, the method may also face noise and biases arising from limited available information or other unaccounted effects. To evaluate approximately the accuracy of the method, we can perform a consistency check by splitting the acquisitions of each spectrometer into two equal parts (splitting between consecutive swath or dwell) and perform a relative SRF retrieval per spectrometer with itself. As we expect to retrieve only identity matrices, one can evaluate potential biases and noises by analysis of the average and standard deviations of the results.

## V. Conclusion

In conclusion, we have demonstrated the existence of a mathematical solution allowing to retrieve a relative SRF between two spectrometers and harmonize their acquisitions, from a dataset of collocated acquisitions. The solution is made robust on purpose the presence of collocation errors and radiometric noise; therefore, it is easily applicable to instruments composed of several spectrometers whose footprints are not even spatially overlapping. The method was already proved effective to characterize Metop-IASI [2] and harmonize MTG-I FCI [3], we are looking now to new fields of applications.